\documentclass[preprint,showpacs,preprintnumbers,amsmath,amssymb]{revtex4}
\usepackage{graphicx}
\usepackage{dcolumn}
\usepackage{bm}
\usepackage{epsfig}
\begin{document}
\title{Quark Confinement in Restricted SU(2) Gauge Theory}
\author{S. Deldar}\altaffiliation{Corresponding author: sdeldar@ut.ac.ir}
\author{A. Mohamadnejad}\altaffiliation{a.mohamadnejad@ut.ac.ir}
\affiliation{Department of Physics, University of Tehran, P.O. Box 14395/547, Tehran 1439955961,
Iran}
\begin{abstract}
We apply Zwanziger formalism to Cho restricted $ SU(2) $ theory to obtain the potential in a static quark-antiquark pair.
Cho restricted theory is a self-consistent subset of a non-Abelian $ SU(2) $ gauge theory which tries to describe the infrared regime of Yang-Mills gauge theories.
In Zwanziger formalism, a local Lagrangian depending on two electric and magnetic gauge fields is constructed for the theories where both electric and magnetic charges exist.
Based on this local Lagrangian the propagator and then the potential between quarks is
calculated in two limits: $ m_{C} r \ll 1 $ and $ m_{C} r \gg 1$, where $ m_{C} $ is the mass of the dual gauge
boson and $ r $ is the distance between the quark and the antiquark.
\end{abstract}
\pacs{11.15.Ha, 12.38.Aw, 12.38.Lg, 12.39.Pn} \maketitle

\section{\label{sec:level1}Introduction}

One of the most interesting problems in particle physics is the confinement of quarks
in quantum chromodynamics (QCD). There are various frameworks to solve this difficult problem. 
One is the dual superconductor picture proposed by Nambu, t' Hooft, Mandelstam, and others in the 1970s. which is based on the existence of magnetic monopoles in QCD and their condensations \cite{Nambu}.
Monopole condensation can explain the confinement via the dual Meissner effect. Indeed, one can easily argue that the dual Meissner effect guarantees the confinement as a consequence of monopole condensation. However there has not been a satisfactory proof for monopole condensation
in QCD, yet.

If the dual superconductivity is considered as a promising mechanism for describing quark confinement, the existence of magnetic monopoles should be confirmed. The quantum theory of the magnetic monopole, initially proposed by Dirac \cite{Dirac}, has played a very important role in QED. Dirac showed that the existence of magnetic monopoles leads to the quantization of electric charges. In the meantime, it was discovered that magnetic monopoles 
would occur naturally in non-Abelian models like the Georgi-Glashow model but without the difficulties of the  Dirac magnetic monopoles.
In this model the non-Abelian local symmetry is broken into an electromagnetic $ U(1) $ symmetry by the Higgs mechanism \cite{'t Hooft}.
These magnetic monopoles called 't Hooft-Polyakov monopoles, are topological solitons which
are massive and cannot become superconducting in order to explain confinement. On the other hand, in the pure Yang-Mills theory 
no matter fields exist. In the absence of matter fields, two popular methods have been introduced to
extract magnetic monopole degrees of freedom in the Yang-Mills theory. One is the Abelian projection, which is a partial gauge fixing proposed by 't Hooft \cite{Abelian Projection}.
The second is a field decomposition method where new variables are introduced by Cho, Faddeev and Niemi \cite{Cho,Faddeev,Shabanov}.
The first method leads to Abelian dominance \cite{Yotsuyanagi} and magnetic
monopole dominance \cite{Stack}  in the maximal Abelian gauge \cite{Kronfeld}. The second method enables one to establish the dual superconductivity in Yang-Mills theory. By applying the second method, one is able to extract the Abelian part of the theory that confines the quark. 

In this paper we use the field decomposition method proposed by Cho in his paper \cite{Cho}.
In this method, an extra symmetry called magnetic symmetry 
decreases the dynamical degrees of freedom. It restricts the original gauge theory and makes it Abelian.
In fact, the original $ SU(2) $ gauge field is decomposed to two fields, electric and magnetic, and the Lagrangian
is rewritten based on these two new fields. Although the two potentials appear in a symmetric way in the Lagrangian, there still exists a significant
disparity between them. While the electric potential is regular, the magnetic
one is singular and it contains a string singularity. In addition, the magnetic potential that describes monopoles is a "spacelike"
potential while the electric one describes isocharges with a "timelike" potential. Cho tried to solve these apparent asymmetries
by introducing the concept of the dual magnetic potentials. But it leads to some singularities in both electric and magnetic potentials.
We remove these singularities to obtain a local Lagrangian with regular electric and magnetic
potentials by applying Zwanziger formalism \cite{Zwanziger} to the Cho Lagrangian.
As a result, we obtain a dual of Ginsburg-Landau Lagrangian, which can describe quark confinement. Using this Lagrangian,
 we calculate the potential in a static quark-antiquark pair 
in two different limits: $ m_{C} r \gg 1$  and $ m_{C} r \ll 1 $, where $ m_{C} $ is the mass of the dual gauge
boson and $ r $ is the distance between  quarks.
 The limit $ m_{C} r \gg 1$ has already been discussed with different approaches \cite{Suganuma, Chernodub, Ripka}.

In the next section, we briefly review the Cho decomposition method as a restricted gauge theory.
In Sec. III, we apply Zwanziger formalism and improve the Cho Lagrangian to a dual Ginsburg-Landau Lagrangian in the framework of a dual-superconductor picture.
In Sec. IV, the Coulombic and the linear parts of the potential are obtained by calculating the propagator from the Lagrangian in two different limits.
Finally, the conclusion and summary are given in Sec. V.

\section{Cho Decomposition Method}

In Cho formalism, an extra symmetry called magnetic symmetry is applied to the theory by a unit vector field $ \hat{m} $ in the adjoint representation
\begin{equation}
D_{\mu}\hat{m}={\partial}_{\mu}\hat{m}+g\overrightarrow{B}_{\mu}\times\hat{m}=0,
\end{equation}
where $ \overrightarrow{B}_{\mu} $ is an $ SU(2) $ gauge potential.
Equation (1) can be solved exactly for $ \overrightarrow{B}_{\mu} $
\begin{equation}
\overrightarrow{B}_{\mu}=A_{\mu}\hat{m}-\frac{1}{g}\hat{m}\times{\partial}_{\mu}\hat{m},
\end{equation}
where $ A_{\mu} $ is the Abelian part of $ \overrightarrow{B}_{\mu} $ that is not restricted by Eq. (1).
The unrestricted part $ A_{\mu} $ is called the electric potential and the other part, which is restricted, is called the magnetic potential.

Using $ \overrightarrow{B}_{\mu} $ of Eq. (2), one can easily show that the field strength is decomposed to
two parts $ F_{\mu\nu} $ and $ H_{\mu\nu} $

\begin{eqnarray}
\overrightarrow{G}_{\mu\nu}=\partial_{\mu}\overrightarrow{B}_{\nu}-\partial_{\nu}\overrightarrow{B}_{\mu}+g\overrightarrow{B}_{\mu}\times\overrightarrow{B}_{\nu} \nonumber\\ &&
\hspace{-58mm} =(F_{\mu\nu}+H_{\mu\nu})\hat{m},
\end{eqnarray}
where
\begin{eqnarray}
F_{\mu\nu}=\partial_{\mu}A_{\nu}-\partial_{\nu} A_{\mu}, & & \nonumber\\ &&
\hspace{-40mm} H_{\mu\nu}=-\frac{1}{g} \hat{m} \,.\, (\partial_{\mu} \hat{m}\times\partial_{\nu} \hat{m}).
\end{eqnarray}
Equation (3) shows that $ \overrightarrow{G}_{\mu\nu} $ is parallel to $ \hat{m} $, and it is made of two parts:
$ F_{\mu\nu} $, which comes from the unrestricted potential $ A_{\mu} $, and $ H_{\mu\nu} $, which comes from the restricted part which contains $ \hat{m} $. It is natural to call $ F_{\mu\nu} $ the electric strength and $ H_{\mu\nu} $ the magnetic strength.

It is possible to associate a magnetic potential $ C^{*}_{\mu} $ to the field strength $ H_{\mu\nu} $ by choosing a hedgehog configuration for $ \hat{m} $
\begin{equation}
\hat{m}=\frac{r^{a}}{r}=
\begin{pmatrix}
\sin{\alpha} \, cos{\beta} \\
\sin{\alpha} \, sin{\beta} \\
\thickspace cos{\alpha}
\end{pmatrix}.
\end{equation}
Using Eq. (5) in Eq. (4), $ H_{\mu\nu} $ is
\begin{eqnarray}
H_{\mu\nu}=\partial_{\mu} C^{*}_{\nu} - \partial_{\nu} C^{*}_{\mu},
\end{eqnarray}
where
\begin{equation}
C^{*}_{\mu}=\frac{1}{g} cos{\alpha} \, \partial_{\mu}\beta.
\end{equation}
$ C^{*}_{\mu} $ is called magnetic potential.

Now, an $ SU(2) $ QCD Lagrangian is constructed with $ {G}_{\mu\nu} $ defined in Eq. (3).
Fermions are included as well

\begin{equation}
L=-\frac{1}{4} F_{\mu\nu}^{2}-\frac{1}{2} F_{\mu\nu} H_{\mu\nu}-\frac{1}{4} H_{\mu\nu}^{2} + \overline{\Psi}(i\gamma^{\mu}D_{\mu}-m)\Psi.
\end{equation}
Fixing the gauge by choosing $ \hat{m} $ along the third axis, $ B_{\mu} $ is obtained
\begin{equation}
B_{\mu}=(A_{\mu}+C^{*}_{\mu}) \, \frac{1}{2}\sigma_{3}.
\end{equation}
This magnetic gauge is chosen to make the SU(2) Lagrangian Abelian. After breaking the symmetry to the U(1) gauge group, $ B_{\mu} $ will be an Abelian gauge field. However, unlike QED, a magnetic current $ k_{\nu} $emerges in the theory 
\begin{equation}
\partial^{\mu} G_{\mu\nu}^{*} = \partial^{\mu} H_{\mu\nu}^{*} = k_{\nu} \neq 0,
\end{equation}
where $ G_{\mu\nu}^{*} $ is the dual field strength, $ H_{\mu\nu}^{*} $ is the dual magnetic field strength and $ k_{\nu} $ is the magnetic current four-vector. Monopole current results from the magnetic potential $ C_{\mu}^{*} $ and it appears because of some
unusual magnetic potential features. This potential is spacelike and contains a Dirac string.
Actually, one can obtain Wu-Yang monopoles from $ C_{\mu}^{*} $ by choosing the appropriate $ \alpha $ and $ \beta $ in Eq. (7)\cite{Cho}. Since in field theory we have a field for every particle, we must introduce a field for magnetic monopoles in the Lagrangian (8). 
Moreover, $ C_{\mu}^{*} $ is spacelike and this is not desirable. In addition, the Dirac string that appeared
in the theory is an unphysical singularity. For a field-theoretic description, it is necessary to remove these undesirable features. 
Adding a complex scalar field $ \phi $ for the monopole, one has to define a dual gauge field $ B_{\mu}^{*} $ to couple to the monopole field as well \cite{Cho}

\begin{equation}
L=-\frac{1}{4} F_{\mu\nu}^{2}-\frac{1}{2} F_{\mu\nu} H_{\mu\nu}-\frac{1}{4} {H_{\mu\nu}^{*}}^{2} + \overline{\Psi}(i\gamma^{\mu}D_{\mu}-m)\Psi
+ \vert (\partial_{\mu}+i \frac{4\pi}{g} B_{\mu}^{*}) \phi \vert^{2} - V(\phi^{*}\phi).
\end{equation}
However, there still exist spacelike potentials $ A_{\mu}^{*} $ and $ C_{\mu}^{*} $ in the Lagrangian. These potentials contain Dirac strings.
In the next section we remove these potentials by Zwanziger formalism \cite{Zwanziger} so that a final
local Lagrangian is obtained, and it depends on matter fields that are spinor fields for quarks, scalar fields for monopoles,
and regular timelike electric and magnetic potentials.

\section{Applying Zwanziger Formalism to Cho Restricted theory}

A local Lagrangian that contains electric and magnetic charges and leads to local field equations without unphysical singularities like Dirac strings has been introduced by Zwanziger \cite{Zwanziger}. It depends on two electric and magnetic four-potentials, and an arbitrary fixed four-vector. Using this framework we can avoid the undesirable features of Lagrangian of Eq. (11) by constructing a Lagrangian depending on a spinor field $ \Psi $ for a quark, a scalar field $ \Phi $ for a monopole, an electric potential $ A_{\mu} $, a magnetic potential $ C_{\mu} $, and a fixed spacelike four-vector $ n_{\mu} $.

In Cho formulation
\begin{eqnarray}
\partial_{\mu} G^{\mu\nu} = j^{\nu}, & & \nonumber\\ &&
\hspace{-25mm} \partial_{\mu} {G^{*}}^{\mu\nu} = k^{\nu},
\end{eqnarray}
where $ j^{\nu} $ is an electric current.

In Zwanziger formalism both electric and magnetic currents are Noether currents. The general solutions for the above equations are \cite{Zwanziger}
\begin{eqnarray}
G= ({\partial}\wedge{A})-(n.{\partial})^{-1}({n}\wedge{k})^{*}, & & \nonumber\\ &&
\hspace{-60mm} G^{*}= ({\partial}\wedge{C})+(n.{\partial})^{-1}({n}\wedge{j})^{*}. 
\end{eqnarray}
G can be described locally in terms of the electric and magnetic potentials $ A $ and $ C $ \cite{Zwanziger}.
Since for a field theoretical description, we need local fields without singularities, we have been motivated to apply Zwanziger formalism to the Cho-restricted theory.
In Zwanziger formalism, there exist electric and magnetic potentials, plus a spinor field for a fermionic particle that has both electric and magnetic charges.
To get the same physics as the Cho-restricted theory but without singularities, we add a scalar monopole field $ \phi $ to the Zwanziger Lagrangian that couples to the magnetic potential $ C $. In addition, we couple the fermionic field $ \psi $ to the electric potential $ A $. In contrast to Zwanziger formalism where $ \psi $ has electric and magnetic charges, we associate the electric charge to $ \psi $ and the magnetic charge to $ \phi $. The final local Lagrangian is
\begin{equation}
\begin{split}
L&=-\frac{1}{2n^{2}} [n.({\partial}\wedge{A})]^{\nu}[n.({\partial}\wedge{C})^{*}]_{\nu}+\frac{1}{2n^{2}} [n.({\partial}\wedge{C})]^{\nu}[n.({\partial}\wedge{A})^{*}]_{\nu} \\
 & \quad -\frac{1}{2n^{2}} [n.({\partial}\wedge{A})]^{2}-\frac{1}{2n^{2}} [n.({\partial}\wedge{C})]^{2} +\overline{\Psi}(i\gamma_{\mu}\partial^{\mu}-g\gamma_{\mu}A^{\mu}\tau_{3}-m)\Psi \\
 & \quad + \vert (\partial_{\mu}+i \frac{4\pi}{g} C_{\mu}) \phi \vert^{2} - V(\phi^{*}\phi),
\end{split}
\end{equation}
where
\begin{equation*}
V(\phi^{*}\phi)=M^{2}\phi^{*}\phi+\lambda(\phi^{*}\phi)^{2}.
\end{equation*}
Choosing the above potential, the spontaneous symmetry breaking would be possible and one can discuss the condensation of monopoles and confinement.
In the absence of a gauge field, the vacuum is at
\begin{equation*}
\vert\phi\vert = a = (\frac{-M^2}{2\lambda})^{\frac{1}{2}}.
\end{equation*}
Expanding $ \phi $ around this vacuum
\begin{equation*}
\phi = a + \frac{\phi_{1}+ i \phi_{2}}{\sqrt{2}}
\end{equation*}
the Lagrangian is obtained in terms of the physical fields $ A $, $ C $, $ \psi $, and $ \phi_{1} $
\begin{equation}
\begin{split}
L&=-\frac{1}{2n^{2}} [n.({\partial}\wedge{A})]^{\nu}[n.({\partial}\wedge{C})^{*}]_{\nu}+\frac{1}{2n^{2}} [n.({\partial}\wedge{C})]^{\nu}[n.({\partial}\wedge{A})^{*}]_{\nu} \\
 & \quad -\frac{1}{2n^{2}} [n.({\partial}\wedge{A})]^{2}-\frac{1}{2n^{2}} [n.({\partial}\wedge{C})]^{2} +\overline{\Psi}(i\gamma_{\mu}\partial^{\mu}-g\gamma_{\mu}A^{\mu}\tau_{3}-m)\Psi \\
 & \quad + \frac{1}{2} m_{C}^{2} C_{\mu}C^{\mu} + \frac{1}{2} (\partial_{\mu}\phi_{1})^2-\frac{1}{2}m_{\phi}^{2} \phi_{1}^{2} + coupling \, \, terms,
\end{split}
\end{equation}
where $ m_{C}^{2}=\frac{-M^{2}}{\lambda} (\frac{4\pi}{g})^{2} $ and $ m_{\phi}^{2} = -2M^{2} $. After approximating the monopole field as the constant mean field that exterminates
$ V(\phi^{*}\phi) $, we get the final Lagrangian
\begin{equation}
\begin{split}
L&=-\frac{1}{2n^{2}} [n.({\partial}\wedge{A})]^{\nu}[n.({\partial}\wedge{C})^{*}]_{\nu}+\frac{1}{2n^{2}} [n.({\partial}\wedge{C})]^{\nu}[n.({\partial}\wedge{A})^{*}]_{\nu} \\
 & \quad -\frac{1}{2n^{2}} [n.({\partial}\wedge{A})]^{2}-\frac{1}{2n^{2}} [n.({\partial}\wedge{C})]^{2} +\overline{\Psi}(i\gamma_{\mu}\partial^{\mu}-g\gamma_{\mu}A^{\mu}\tau_{3}-m)\Psi \\
 & \quad +\frac{1}{2} m_{C}^{2} C_{\mu}^{2}.
\end{split}
\end{equation}

In the next section, by using the above Lagrangian we study the interquark potential by the gluon propagator obtained from the nonperturbative sector.

\section{Quark Confinement Potential}

The static potential between a heavy quark-antiquark pair can be obtained from the energy of the vacuum where the static quark and antiquark exist. Information on confinement is included in the gluon propagator, which
gives the strong interaction in the infrared sector. The vacuum energy $ V (j) $ in the presence of the static quark sources j is obtained
from \cite{Suganuma}
\begin{equation}
Z = \langle0\vert{e^{i\int(L + j_{\mu}A^{\mu}) d^{4}x}}\vert0\rangle = N \int DA_{\mu} DC_{\mu} e^{i \int (L + j_{\mu} A^{\mu}) d^{4} x} = e^{-i V(j) T}.
\end{equation}
Integrating out with respect to the dual gauge field $ C_{\mu} $, the Lagrangian becomes
\begin{equation}
L = - \frac{1}{4} F_{\mu\nu} F^{\mu\nu} + \frac{1}{2} A^{\mu} K_{\mu\nu} A^{\nu} + \overline{\Psi}(i\gamma_{\mu}\partial^{\mu}-g\gamma_{\mu}A^{\mu}\tau_{3}-m)\Psi,
\end{equation}
where
\begin{eqnarray}
K^{\mu\nu} \equiv \frac{n^{2} m_{C}^{2}}{(n.\partial)^{2} + n^{2} m_{C}^{2}} X^{\mu\nu}, & & \nonumber\\ &&
\hspace{-51mm} X^{\mu\nu} \equiv \frac{1}{n^{2}} \epsilon_{\lambda}^{\mu\alpha\beta} \epsilon^{\lambda\nu\gamma\delta} n_{\alpha} n_{\gamma} \, \partial_{\beta} \, \partial_{\delta}.
\end{eqnarray}
A quench approximation is used to remove the quantum effects of the dynamical quarks. The external source j is introduced to represent heavy quarks. Integrating out $ A_{\mu} $ in the Lorenz gauge $ \partial_{\mu} A^{\mu} = 0 $ and $ L_{GF}=-\frac{1}{2\alpha_{g}}  (\partial_{\mu} A^{\mu})^{2} $, the nonlocal current-current correlation is obtained
\begin{equation}
L_{j} = - \frac{1}{2} j_{\mu} D^{\mu\nu} j_{\nu},
\end{equation}
where $ D^{\mu\nu} $ is the propagator of the diagonal gluons
\begin{equation}
D_{\mu\nu} = \frac{1}{\partial^{2}} [ g_{\mu\nu} + (\alpha_{g} -1) \frac{\partial_{\mu} \, \partial_{\nu}}{\partial^{2}}] - \frac{1}{\partial^{2}} \frac{m_{C}^{2}}{\partial^{2} + m_{C}^{2}} \frac{n^{2}}{(n.\partial)^{2}} X_{\mu\nu}.
\end{equation}
The nonperturbative effect is included in the second term. The action is
\begin{eqnarray}
S_{j} \equiv \int d^{4}x L_{j} & & \nonumber\\ &&
\hspace{-25mm} = \int \frac{d^{4}k}{(2\pi)^{4}} \frac{1}{2} j^{\mu} (-k) [ \frac{1}{k^{2} - m_{C}^{2}} g_{\mu\nu} + \frac{- m_{C}^{2}}{k^{2} - m_{C}^{2}} \frac{n^{2}}{(n.k)^{2}} (g_{\mu\nu} - \frac{n_{\mu} n_{\nu}}{n^{2}}) ] j^{\nu} (k),
\end{eqnarray} 
where $ j_{\mu}(k) $ is the Fourier component of $ j_{\mu}(x) $.

Now consider a system of a heavy quark-antiquark pair with opposite color charges located at $ \textbf{a} $ and $ \textbf{b} $, respectively.
The quark current is given by
\begin{equation*}
\begin{split}
j_{\mu}(x) & =Q \, g_{\mu0} [\delta^{3}(\textbf{x}-\textbf{b}) - \delta^{3}(\textbf{x}-\textbf{a})], \\
j_{\mu}(k)& =Q \, g_{\mu0} \, 2\pi \delta(k_{0}) (e^{-i \textbf{k}.\textbf{b}} - e^{-i \textbf{k}.\textbf{a}}).
\end{split}
\end{equation*}
Therefore, $ S_{j} $ of Eq. (22) becomes
\begin{equation}
S_{j} = - Q^{2} \int dt \int \frac{d^{3}k}{(2\pi)^{3}} \frac{1}{2} (1-e^{i\textbf{k}.\textbf{r}})(1-e^{-i\textbf{k}.\textbf{r}}) [\frac{1}{\textbf{k}^{2} + m_{C}^{2}} + \frac{m_{C}^{2}}{\textbf{k}^{2} + m_{C}^{2}} \frac{1}{(\textbf{n}.\textbf{k})^{2}}],
\end{equation}
$ \textbf{n} $ is a unit vector and $ \textbf{r}=\textbf{b}-\textbf{a} $ is a vector which connects the quark to the antiquark.
We choose $ \textbf{n} $ parallel to $ \textbf{r} $ but $ \textbf{n} $ may be chosen in a different direction \cite{Suzuki}.
After subtracting the contribution of the self-energy of the quark and antiquark we get  the static quark-antiquark potential which is written in two parts
\begin{equation}
\begin{split}
V(r) &=V_{Yukawa}(r) + V_{Linear} (r), \\
V_{Yukawa}(r) &=  -Q^{2} \int \frac{d^{3}k}{(2\pi)^{3}} cos(\textbf{k}.\textbf{r}) \frac{1}{\textbf{k}^{2} + m_{C}^{2}} = \frac{- Q^{2}}{4\pi} \frac{e^{-m_{C}r}}{r}, \\
V_{Linear} (r) &= -Q^{2} \int \frac{d^{3}k}{(2\pi)^{3}} cos{(\textbf{k}.\textbf{r})} \frac{m_{C}^{2}}{\textbf{k}^{2} + m_{C}^{2}} \frac{1}{(\textbf{n}.\textbf{k})^{2}}.
\end{split}
\end{equation}
The integration for $ V_{Yukawa}(r) $ is done easily but an exact calculation cannot be done for $ V_{Linear} (r) $. 
There are some approaches leading to different results for calculating $ V_{Linear} (r) $ \cite{Suganuma,Chernodub,Ripka,Suzuki}. In some of them a cutoff is used to make the
 integral converge at large $ k $ \cite{Chernodub,Ripka}  or $ k_{\bot} $ \cite{Suganuma}, where $ k_{\bot} $ is the perpendicular component of $ \textbf{k} $ with respect to
 $ \textbf{n} $ or $ \textbf{r} $ (Fig. 1). However, Suzuki used no cutoff and $ \textbf{n} $ and $ \textbf{r} $ are no longer parallel in his approach \cite{Suzuki}.
\begin{figure}[]
%\vspace{70pt}
\includegraphics*[width=14cm]{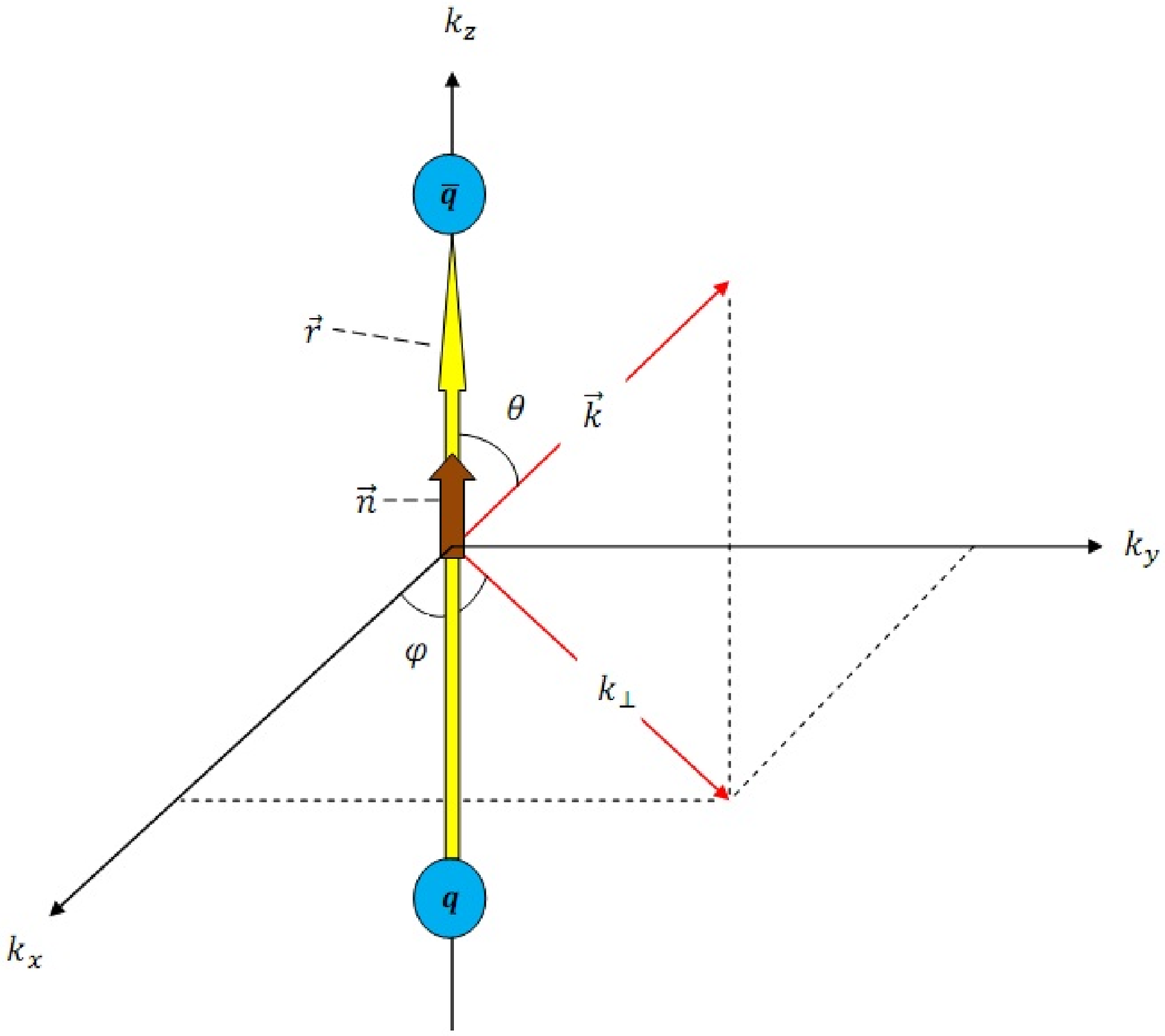}\hspace{14cm}
\caption{The relative situations of vectors $ \textbf{k} $, $ \textbf{r} $, and $ \textbf{n} $} \label{Figure_1}
\end{figure}

In this paper, we calculate the linear term in two limits:
$ m_{C}r \gg 1 $ and $ m_{C}r \ll 1 $. For $ m_{C}r \gg 1 $, $ m_{\phi} $ is used as a cutoff to converge the k integral at large k.
Applying a cutoff for this regime to converge the k integral has been done before but with different methods \cite{Suganuma,Chernodub,Ripka}. For the $ m_{C}r \ll 1 $ limit, we use $ \varepsilon $ as a cutoff for $ cos(\theta) $, where $ \theta $ is the angle between $ \textbf{k} $ and $ \textbf{r} $ as shown in Fig. 1.
This cutoff makes the integral converge at $ cos(\theta) = 0 $. 
In fact, this cutoff makes a constraint on $ \textbf{k} $ for $ m_{C}r \ll 1 $. In this limit, the flux tube between the quark-antiquark pair 
may not be approximated by a thick and very long vortex. Therefore, the situation is not the same as $ m_{C}r \gg 1 $, where one can approximate the thickness of the vortex with the inverse 
of the maximum $ k_{\bot} $, the mass of the monopole. It means that $ \textbf{k} $ cannot be chosen in the direction $ k_{\bot} $, but the angle between $ \textbf{k} $ and $ k_{\bot} $
 should have a small deviation from zero, which is discussed after Eq. (31).

After applying these limits we get

\begin{equation}
V_{Linear} (r) = \frac{Q^{2}m_{C}^{2}}{8\pi} ln{[\frac{m_{C}^{2} +m_{\phi}^{2}}{m_{C}^{2}}]} r
\end{equation}
 for $ m_{C}r \gg 1 $ and
\begin{equation}
V_{Linear} (r) = \frac{Q^{2}m_{C}^{2}}{8\pi} ln[{\varepsilon^{-2}}] r
\end{equation}
for $ m_{C}r \ll 1 $. The details of our calculations are shown in the Appendix.
Comparing the string tensions of Eqs. (25) and (26) obtained from the two limits $ m_{C}r \gg 1 $ and $ m_{C}r \ll 1 $, one can fix 
$ \varepsilon $ versus the physical quantities $ m_{C} $ and $ m_{\phi} $

\begin{equation}
\varepsilon=cos{\theta_{c}}=\frac{m_{C}}{\sqrt{m_{C}^2+m_{\phi}^2}}.
\end{equation}
If $ m_{\phi} \rightarrow \infty $, we get $ \varepsilon=cos{\theta_{c}} \rightarrow 0 $, which means that no cutoff is used.

Finally, we get the quark-antiquark potential
\begin{equation}
V(r) =  \frac{- Q^{2}}{4\pi} \frac{e^{-m_{C}r}}{r} + \sigma r,
\end{equation}
where $\sigma$ is the string tension of the quark-antiquark pair
\begin{equation}
\sigma = \frac{Q^{2}m_{C}^{2}}{8\pi} ln{\varepsilon^{-2}}.
\end{equation}
The first part of the linear potential represents the Coulombic potential if $ m_{C} \rightarrow 0 $ and the second part shows confinement of the quark-antiquark pair.
\begin{figure}[]
%\vspace{70pt}
\includegraphics*[width=14cm]{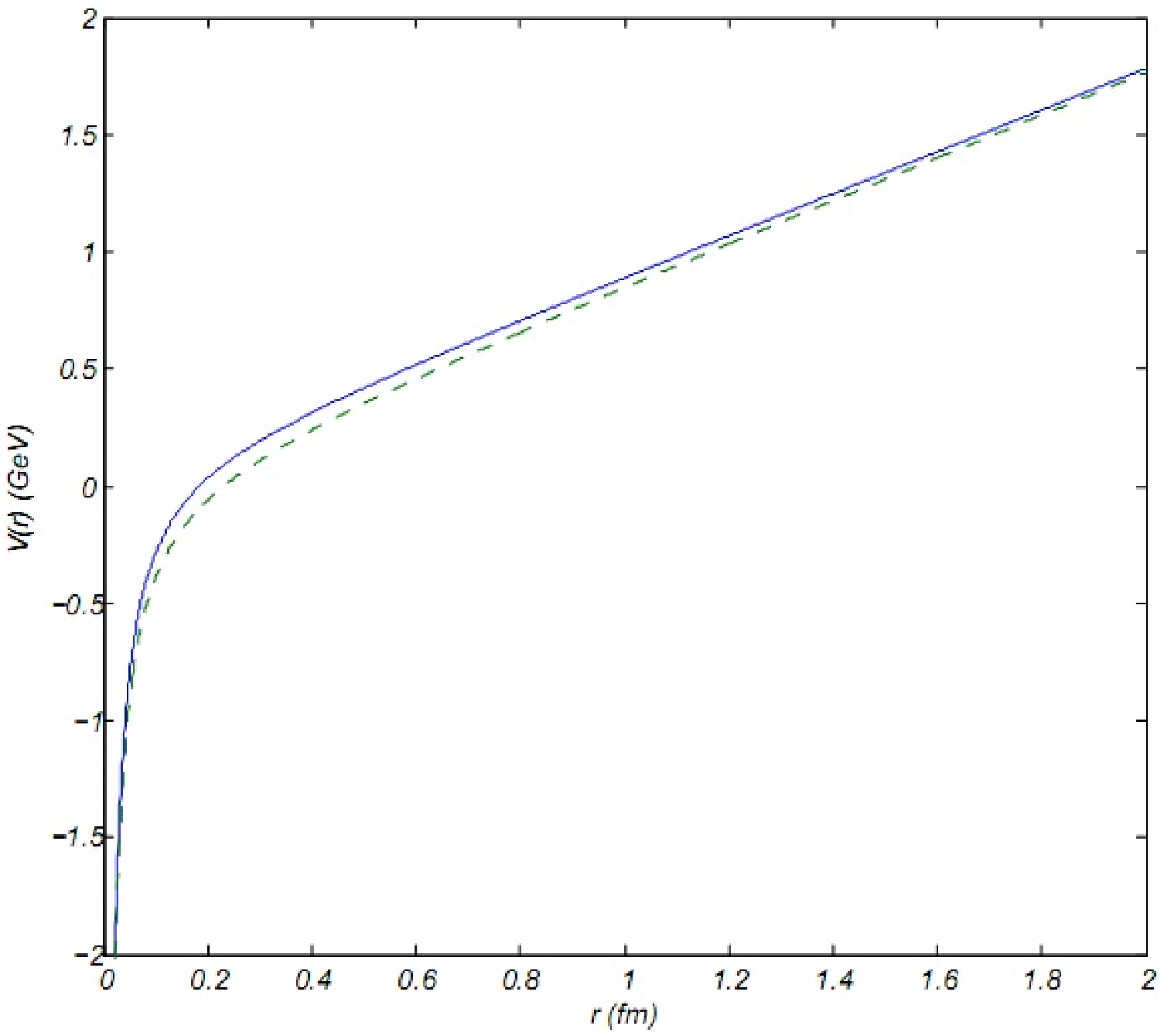}\hspace{14cm}
\caption{The solid curve is obtained by fixing the parameters of the potential calculated in this paper. Dashed curve shows the lattice results.} \label{Figure_2}
\end{figure}

We calculate the values of the parameters of Eqs. (27) and (28) using the data of the Monte Carlo simulations \cite{Huntley}.
The SU(2) static quark-antiquark potential is obtained by
\begin{equation}
V(r) = - \frac{\alpha}{r} + k \, r + constant
\end{equation}
with $ \alpha \thicksim 0.244 $ and $ k = (420 MeV)^{2} $ from the simulation results. We choose
\begin{equation}
Q = \frac{g}{2} = 1.75, \, \, \, , m_{C} = 480 \, MeV,    \, \, \, , m_{\phi} = 11 \, GeV ,
\end{equation}
to reproduce the SU(2) static quark-antiquark potential and we get $ \varepsilon=0.043 $ and $ \theta_{c}=87.8 $ by fixing the above parameters. The result is shown in Fig. 2, where the potential versus distance is plotted for both potentials calculated in this paper and from the lattice data.

\section{Conclusion}

One can restrict the $ SU(2) $ QCD by adding an extra symmetry called magnetic symmetry (Cho decomposition method). As a result of this restriction, magnetic monopoles appear 
in the theory. However, the Cho Lagrangian has some problems like having a Dirac string and spacelike potentials. We solve these undesirable problems by the Zwanziger dual variables method. 
We obtain a Lagrangian that is a dual of the Ginsburg-Landau Lagrangian such that a symmetry breaking can occur and monopole condensation makes the magnetic potential $ C_{\mu} $
 massive. This massive magnetic potential changes the usual propagator to a propagator that leads to a linear potential known as the quark-confinement potential in a quark-antiquark pair.
The string tension is calculated for two different limits, $ m_{C}r \gg 1 $ and $ m_{C}r \ll 1 $,  by making physical constraints on $\textbf{k}$. 
One can generalize this method for the other SU(N) gauge groups. Other decomposition methods may be investigated to obtain confinement potential as well.

\section{\boldmath Acknowledgments}

We are grateful to the research council of University of Tehran for supporting this study.

\begin{center}
\maketitle{\textbf{Appendix}}
\end{center}

We find the solution of the integral of Eq. (24). First, the Yukawa potential is obtained from the following formula:
\begin{equation*}
\begin{split}
V_{Yukawa}(r) & =  -Q^{2} \int \frac{d^{3}k}{(2\pi)^{3}} cos(\textbf{k}.\textbf{r}) \frac{1}{\textbf{k}^{2} + m_{C}^{2}} \\
& =  -Q^{2} \int_{0}^{\infty} \int_{-1}^{1} \int_{0}^{2\pi} \frac{k^{2} \, dk \, dx \, d\phi}{(2 \pi)^{3}} cos{(krx)} \frac{1}{k^{2} + m_{C}^{2}}.
\end{split}
\end{equation*}
Keeping k fixed and integrating with respect to $ \phi $ and $ x $, where $ x $ is $ cos\theta $ and $ \theta $ is the polar angle in momentum space (Fig. 1), we obtain

\begin{equation*}
\begin{split}
V_{Yukawa}(r) & =  \frac{ -2Q^{2}}{(2 \pi)^{2}r} \int_{0}^{\infty} \frac{ k \, sin{(kr)}}{k^{2} + m_{C}^{2}} \, dk\\
& =  \frac{ -Q^{2}}{(2 \pi)^{2}r} \int_{-\infty}^{\infty} \frac{ y \, sin{y}}{y^{2} + (m_{C}r)^{2}} \, dy.
\end{split}
\end{equation*}
From the calculus of residues
\begin{equation*}
\int_{-\infty}^{\infty} \frac{ y \, sin{y}}{y^{2} + a^{2}} \, dy = \pi \, e^{-a}.
\end{equation*}
Finally, $ V_{Yukawa} $ is
\begin{equation*}
V_{Yukawa}(r) = \frac{- Q^{2}}{4\pi} \frac{e^{-m_{C}r}}{r}.
\end{equation*}

Using the same procedure for the Linear potential, we obtain
\begin{equation*}
\begin{split}
V_{Linear} (r) &= -Q^{2} \int \frac{d^{3}k}{(2\pi)^{3}} cos{(\textbf{k}.\textbf{r})} \frac{m_{C}^{2}}{\textbf{k}^{2} + m_{C}^{2}} \frac{1}{(\textbf{n}.\textbf{k})^{2}} \\
& = -Q^{2} \int_{0}^{\infty} \int_{-1}^{1} \int_{0}^{2\pi} \frac{k^{2} \, dk \, dx \, d\phi}{(2 \pi)^{3}} cos{(krx)} \frac{m_{C}^{2}}{k^{2} + m_{C}^{2}} \frac{1}{k^{2}x^{2}}.
\end{split}
\end{equation*}
Integrating with respect to $ \phi $ and using the $ cos{x}=1-2sin^{2}{\frac{x}{2}} $ formula we have
\begin{equation*}
\begin{split}
V_{Linear} (r) &= \frac{-Q^{2} m_{C}^{2}}{(2\pi)^{2}} \int_{0}^{\infty} \frac{dk}{k^{2} + m_{C}^{2}} \int_{-1}^{1} \frac{1-2sin^{2}({\frac{krx}{2}})}{x^{2}} dx  \\
& = \frac{Q^{2} m_{C}^{2} \, r}{(2\pi)^{2}} \int_{0}^{\infty} \frac{k \, dk}{k^{2} + m_{C}^{2}} \int_{-\frac{kr}{2}}^{\frac{kr}{2}} \frac{sin^{2}{y}}{y^{2}} dy.
\end{split}
\end{equation*}
We neglect the term $ \frac{1}{x^{2}} $ because it is independent of $ r $. For regions where $ m_{C} r \gg 1 $ we can use the following approximation:
\begin{equation*}
\int_{-\frac{kr}{2}}^{\frac{kr}{2}} \frac{sin^{2}{y}}{y^{2}} dy \simeq \int_{-\infty}^{\infty} \frac{sin^{2}{y}}{y^{2}} dy = \pi .
\end{equation*}
Then,
\begin{equation*}
V_{Linear} (r) =  \frac{Q^{2} m_{C}^{2} \, r}{4\pi} \int_{0}^{\infty} \frac{k \, dk}{k^{2} + m_{C}^{2}} = \frac{Q^{2}m_{C}^{2}}{8\pi} ln{[\frac{m_{C}^{2} +m_{\phi}^{2}}{m_{C}^{2}}]} r ,
\end{equation*}
where a sharp cutoff $ m_{\phi} $ was introduced to make the k integral converge at large k. We have gotten the same result for $ V_{Linear} $ as Refs. \cite{Suganuma, Ripka} but with different approximations.

For regions where $ m_{C} r \ll 1 $, we evaluate the integral from the first place with a different method. After integration with respect to $ \phi $, $ V_{Linear} $ is obtained:
\begin{equation*}
\begin{split}
V_{Linear} (r) & = \frac{-Q^{2} m_{C}^{2}}{(2\pi)^{2}} \int_{0}^{\infty} \int_{-1}^{1} \, dk \, dx  \frac{1}{k^{2} + m_{C}^{2}} \frac{cos{(krx)}}{x^{2}}   \\
& = \frac{-Q^{2} m_{C}^{2}}{(2\pi)^{2}} \int_{0}^{1} \frac{dx}{x^{2}} \int_{-\infty}^{\infty} \frac{cos{(krx)}}{k^{2} + m_{C}^{2}} dk .
\end{split}
\end{equation*}
To get the second line we have used the fact that the integrand is even with respect to both $ x $ and $ k $. Now, we 
have to use a physical cutoff  for the $ x $ integral, as mentioned in Sec. IV. As a result of this cutoff, no divergence happens in  the integral for small $ x $. Using $ \varepsilon $ instead of zero in the lower limit of the $ x $ integration, we get
\begin{equation*}
\begin{split}
V_{Linear} (r) & =  \frac{-Q^{2} m_{C}^{2}}{(2\pi)^{2}} \int_{\varepsilon}^{1} \frac{dx}{x^{2}} \int_{-\infty}^{\infty} \frac{cos{(krx)}}{k^{2} + m_{C}^{2}} dk \\
& = \frac{-Q^{2} m_{C}^{2}}{(2\pi)^{2}} r \int_{\varepsilon}^{1} \frac{dx}{x} \int_{-\infty}^{\infty} \frac{cos{y}}{y^{2} + (m_{C}rx)^{2}} dy .
\end{split}
\end{equation*}
Using the calculus of residues
\begin{equation*}
\int_{-\infty}^{\infty} \frac{cos{y}}{y^{2} + a^{2}} dy = \frac{\pi}{a}e^{-a}
\end{equation*}
we get
\begin{equation*}
V_{Linear} (r) = \frac{-Q^{2} m_{C}}{4\pi} \int_{\varepsilon}^{1} \frac{e^{-m_{C}rx}}{x^{2}} dx .
\end{equation*}
In addition, for $ m_{C} r \ll 1 $ we can use the following expansion:
\begin{equation*}
e^{-m_{C}rx}=1- m_{C}rx + O( (m_{C}rx)^{2}) .
\end{equation*}
Therefore,
\begin{equation*}
V_{Linear} (r) = \frac{-Q^{2} m_{C}}{4\pi} \int_{\varepsilon}^{1} \frac{1- m_{C}rx}{x^{2}} dx .
\end{equation*}
We again neglect the term $ \frac{1}{x^{2}} $ because it is independent of $ r $. Finally, $ V_{Linear} $ is
\begin{equation*}
V_{Linear} (r) = \frac{Q^{2} m_{C}}{4\pi} \int_{\varepsilon}^{1} \frac{m_{C}rx}{x^{2}} dx = \frac{Q^{2}m_{C}^{2}}{8\pi} ln[{\varepsilon^{-2}}] r .
\end{equation*}


\begin{thebibliography}{17}
\bibitem{Nambu}
Y. Nambu, \textit{Phys. Rev. D} \textbf{10} 4262 (1974); G. 't Hooft, in \textit{High Energy Physics},
edited by A. Zichichi (Editorice Compositori, Bologna, 1975);
S. Mandelstam, \textit{Phys. Rep.} \textbf{23} 245 (1976).

\bibitem{Dirac}
P. A. M. Dirac, \textit{Proc. R. Soc.} \textbf{A133} 60 (1931).

\bibitem{'t Hooft}
G. 't Hooft, \textit{Nucl. Phys.}\textbf{ B79} 276 (1974); A. M. Polyakov, A.M., \textit{JETP  Lett.} \textbf{20} 194 (1974).

\bibitem{Abelian Projection}
G. 't Hooft, \textit{Nucl. Phys.} \textbf{B190} 455 (1981).

\bibitem{Cho}
Y. M. Cho, \textit{Phys. Rev. D} \textbf{21} 1080 (1980); \textit{Phys. Rev. D} \textbf{23} 2415 (1981).

\bibitem{Faddeev}
L. Faddeev and A. J. Niemi, \textit{Phys. Rev. Lett}. \textbf{82} 1624 (1999);
\textit{Nucl. Phys.} \textbf{B776} 38 (2007).

\bibitem{Shabanov}
S. V. Shabanov, \textit{Phys. Lett.} \textbf{458B} 322 (1999); \textit{Phys. Lett.} \textbf{B463} 263 (1999).

\bibitem{Yotsuyanagi}
T. Suzuki and I. Yotsuyanagi, \textit{Phys. Rev. D} \textbf{42} 4257 (1990).

\bibitem{Stack}
J. D. Stack, S. D. Neiman, and R. Wensley, \textit{Phys. Rev. D} \textbf{50} 3399 (1994);
H. Shiba and T. Suzuki, \textit{Phys. Lett.} \textbf{B333} 461 (1994).

\bibitem{Kronfeld}
A. Kronfeld, M. Laursen, G. Schierholz and U.-J. Wiese, \textit{Phys. Lett.} \textbf{B198} 516 (1987).


\bibitem{Zwanziger}
D. Zwanziger, \textit{Phys. Rev. D} \textbf{3} 880 (1971).

\bibitem{Suganuma}
H. Suganuma, S. Sasaki, H. Toki, \textit{Nucl. Phys.} \textbf{B435} 207 (1995).

\bibitem{Chernodub}
D. A. Komarov and M. N. Chernodub, \textit{JETP Lett.} \textbf{68} 117 (1998).

\bibitem{Ripka}
G. Ripka, arXiv:hep-ph/0310102v2.

\bibitem{Suzuki}
T. Suzuki, \textit{Prog. Theor. Phys.} \textbf{80} 929 (1988).

\bibitem{Huntley}
A. Huntley and C. Michael, \textit{Nucl. Phys.} \textbf{270} 123 (1986).
 
\end{thebibliography}
\end{document}